\documentclass[superscriptaddress,nofootinbib, amsmath,amssymb, aps,prl,portrait,twocolumn]{revtex4-1}
\usepackage{multirow,dcolumn}
\usepackage{xcolor,colortbl}
\usepackage{longtable}
\usepackage{textcomp}
\usepackage{graphicx}
\usepackage{bm}
\usepackage{lineno}
\usepackage{hyperref}
\hypersetup{colorlinks=true, citecolor=blue, urlcolor=blue, linkcolor=blue}

\usepackage{soul}

\newcommand{\nuc}[2]    {$^{#1}$\textrm{#2}} 

\newcommand{\be}        {\begin{equation}}
\newcommand{\ee}        {\end{equation}}

\definecolor{Gray}{gray}{0.85}
\definecolor{LightCyan}{rgb}{0.88,1,1}
\definecolor{BlueTable}{rgb}{0.30,0.58,0.93}
\definecolor{SREblizzardblue}{rgb}{0.9, 0.9, 0.98}
\definecolor{powderblue}{rgb}{0.69, 0.88, 0.9}
\definecolor{turquoiseblue}{rgb}{0.0, 1.0, 0.94}
\definecolor{skyblue}{rgb}{0.53, 0.81, 0.92}
\definecolor{lightskyblue}{rgb}{0.53, 0.81, 0.98}
\definecolor{lightcornflowerblue}{rgb}{0.6, 0.81, 0.93}
\definecolor{SREblizzardbluemist}{rgb}{0.9, 0.9, 0.98}
\definecolor{SREblizzardblue}{rgb}{0.0, 1.0, 1.0}
\definecolor{SREblizzardblue}{rgb}{0.91, 1.0, 1.0}
\definecolor{SREblizzardblue}{rgb}{.8, 0.9, 0.93}

\begin{document}

\newcommand{\LANL}{Los Alamos National Laboratory, Los Alamos, NM 87545, USA}
\newcommand{\Stanford}{Stanford Institute for Theoretical Physics, Stanford University, Stanford, CA 94305, USA}

\title{Dark Matter Constraints from Isomeric $^{\bf 178m}$Hf}

\affiliation{\LANL}
\affiliation{\Stanford}

\author{D. S. M.~Alves}\affiliation{\LANL}
\author{S. R.~Elliott}\email[]{elliotts@lanl.gov}\affiliation{\LANL}
\author{R.~Massarczyk}\affiliation{\LANL}
\author{S. J.~Meijer}\affiliation{\LANL}
\author{H.~Ramani}\affiliation{\Stanford}

\date{\today}

\begin{abstract}
We describe a first measurement of the radiation from a \nuc{178m}{Hf} sample to search for dark matter. The $\gamma$ flux from this sample, possessed by  Los Alamos National Laboratory nuclear chemistry, was measured with a Ge detector at a distance of 4 ft due to its high activity. We search for $\gamma$s that cannot arise from the radioactive decay of \nuc{178m}{Hf}, but might arise from the production of a nuclear state due to the inelastic scattering with dark matter. The limits obtained on this $\gamma$ flux are then translated into constraints on the parameter space of inelastic dark matter. Finally, we describe the potential reach of future studies with \nuc{178m}{Hf}.

\end{abstract}

\maketitle

There is irrefutable evidence for the existence of dark matter arising purely from gravitational interactions. Understanding its particle nature is one of the burning questions of 21\textsuperscript{st} century particle physics.
Dark matter candidates at the weak scale arise naturally in theories beyond the Standard Model, such as supersymmetry. Furthermore, weak scale massive particles with weak scale cross-sections---the so-called ``weakly interacting massive particles," or WIMPS---are produced with the correct relic abundance when freezing out from thermal equilibrium in the early universe (WIMP miracle). Direct, indirect, and collider searches for WIMPs have reported repeated null results, setting stringent limits on their model space. 
A review of the present status of the search for dark matter can be found in \cite{snowmassDMReview}.
The derived limits are very restrictive and the lack of an observation has motivated dark matter considerations beyond the classic WIMP description. 

Among other alternative dark matter models are inelastic dark matter (iDM)~\cite{PhysRevD.64.043502,PhysRevD.82.075019,PhysRevD.89.055008,PhysRevD.94.115026}  and strongly interacting dark matter (SIDM)~\cite{PhysRevD.97.115006,PhysRevD.97.115024,Neufeld_2018}. While these models retain the salient features of thermal freeze-out, constraints on their parameter spaces are much less stringent because the threshold energy required for a detectable dark matter-nucleus scattering event is often unavailable---in the case of iDM, due to the large inelastic splitting; and in the case of SIDM, due to the loss of dark matter kinetic energy from its interactions with the overburden rock above deep underground experiments. More specifically, in iDM models, dark matter-nucleus elastic scattering is suppressed, and the dominant scattering process requires an internal transition to an excited dark matter state. If the transition energy is greater than the available kinetic energy of the dark matter-nucleus system, this process is completely shut-off, severely reducing the sensitivity to the iDM parameter space of present experiments focused on WIMPs. In the case of SIDM models, the large dark matter nuclear cross section causes dark matter particles to thermalize through interactions with the Earth, resulting in a velocity too low to produce a measurable interaction by the time they reach the detector's location deep underground. In particular, large scattering cross sections (much higher than electroweak-strength) are still viable for significant swaths of parameter space of both SIDM and iDM models. While this is difficult to obtain with perturbative models of new physics at the TeV scale, composite dark matter models from a strongly interacting dark sector can naturally accommodate these properties \cite{Nussinov:1985xr,Bagnasco:1993st,Alves:2009nf,Lisanti:2009am,SpierMoreiraAlves:2010err}.

In this letter, we consider a recent a proposal to use nuclear metastable states as exothermic-reaction targets for dark matter searches~\cite{Pospelov2020}. The isomer \nuc{178m}{Hf} was suggested as a potential target and its large reservoir of available energy  for transference to a dark matter particle makes it an intriguing candidate. Unfortunately, due to the relatively short half-life (31 y) of this manmade isotope, a large number of target atoms also means significant radioactivity. In this article, we describe a first study of the $\gamma$ spectrum of \nuc{178m}{Hf} to derive first limits on dark matter interactions with this isotope.

\begin{figure}[thp]
 \centering
 \includegraphics[width=0.8\columnwidth]{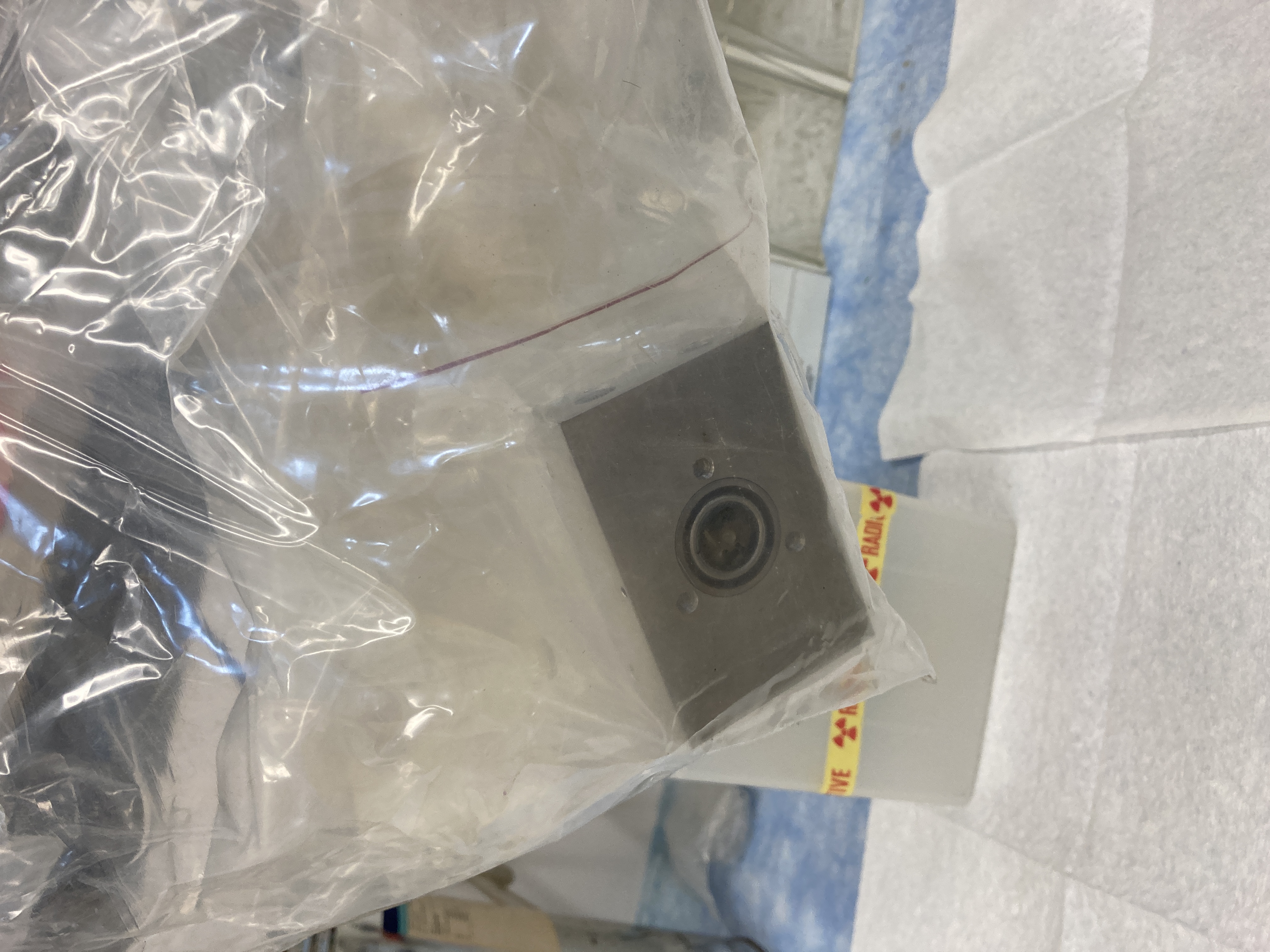}
 \caption{A photo of the \nuc{178m}{Hf} source.}
 \label{fig:HfSource}
\end{figure}

The \nuc{178m}{Hf} sample used in this study was fabricated by the Los Alamos National Laboratory (LANL) nuclear chemistry division  to produce \nuc{172}{Hf} for a \nuc{172}{Hf}/\nuc{172}{Lu} medical generator~\cite{Taylor1998}. Later, this material was used to study the possibility of energy storage in nuclear isomers, specifically \nuc{178m}{Hf}. Reports of triggered isomer decay in \nuc{178m}{Hf} led to a further study~\cite{Ahmad2005} finding no evidence for the effect. The Hafnium sample used for that experiment, and for this measurement, was extracted from a Ta target at the LAMPF accelerator at LANL~\cite{Ahmad2001}. The sample studied here is the \textit{second set} as described in \cite{Ahmad2005} and is shown in Fig.~\ref{fig:HfSource}. It is now relatively old and Hf isotopes other than 178m have decayed away, leaving a rather pure sample.

\begin{figure*}[thp]
 \centering
 \includegraphics[width=0.8\textwidth]{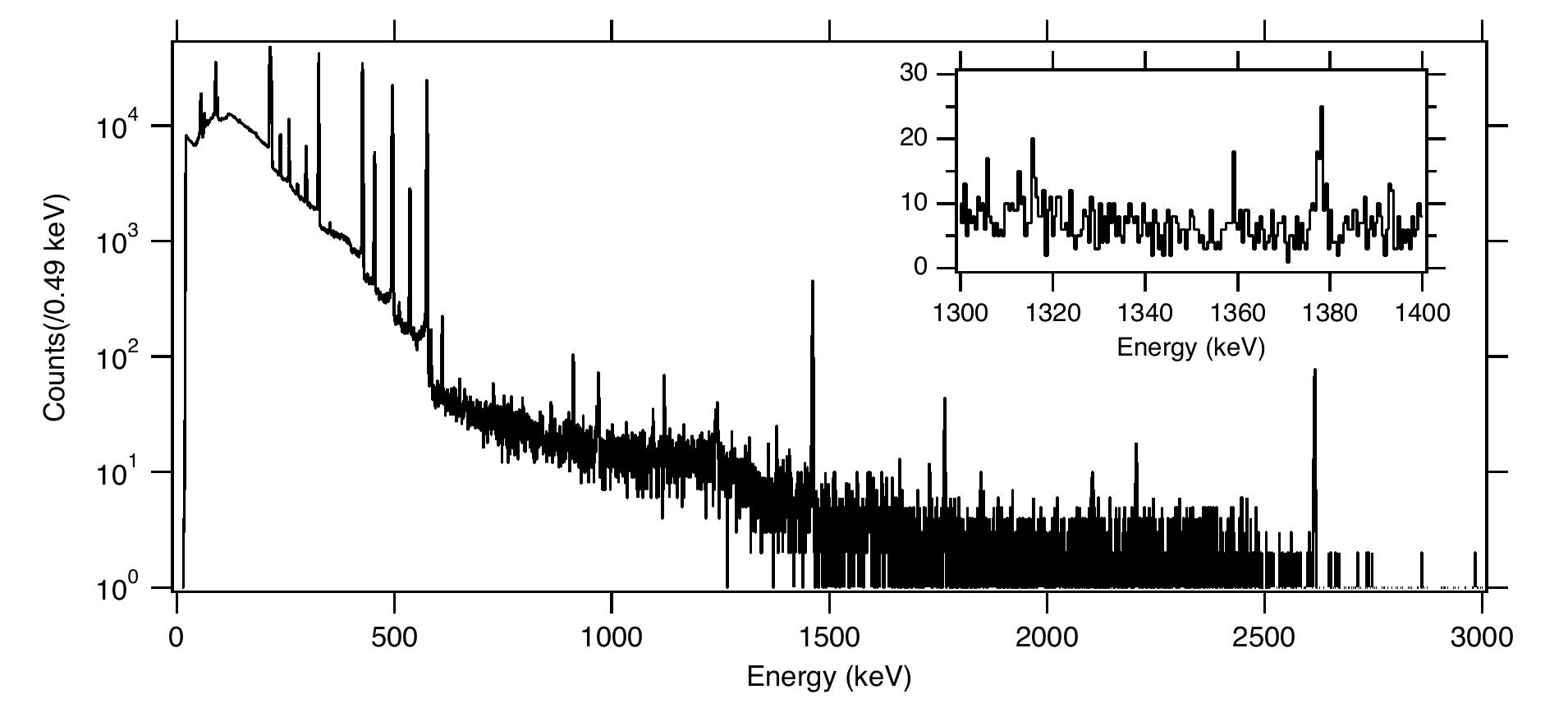}
 \caption{The observed spectrum from the \nuc{178m}{Hf} sample for a live time of 974.2~s. The inset shows the spectrum surrounding 1330~keV. A dark matter-induced $\gamma$ with this energy proved to be the most sensitive test.}
 \label{fig:spectrum}
\end{figure*}

Due to the sample's high activity (20-25~mR/hr on contact), the detector was placed at a distance of 1.2 m (4~ft) in order to minimize dead time from pile-up rejection. The detector was an ORTEC\textsuperscript{\textregistered} Detective-X Ge detector~\cite{ORTEC}. The spectrum, shown in Fig.~\ref{fig:spectrum}, indicates some natural room background lines from the U/Th/K decay chains along with the known \nuc{178m}{Hf} lines. A list of the identified lines are given in Table~\ref{tab:LineIDs}. Note that above 600~keV, the spectrum is dominated by the natural background. Below 600~keV, the spectrum is dominated by \nuc{178m}{Hf} emission.

\begin{table}[htp]
\caption{The stronger lines observed in the spectrum.}
\label{tab:LineIDs}
\centering
\begin{tabular}{cc||cc}
\hline
$E_\gamma$ [keV]	&	Origin					& $E_\gamma$ [keV]			&	Origin	\\
\hline\hline
55.3		&	\nuc{178m}{Hf} x-rays	&	725.9		&	\nuc{212}{Bi}\\
62.7		&	\nuc{178m}{Hf} x-ray  		&	768.4		&	\nuc{214}{Bi} 	\\
68.6		&	Au x-ray  det. mat.			&	794.9		&	\nuc{228}{Ac} 	\\
77.9		&	Au x-ray det. mat.			&	860.6		&	\nuc{208}{Tl}\\
88.6		&	\nuc{178m}{Hf}				&		910.5	&	\nuc{228}{Ac}	\\
93.2		&	 \nuc{178m}{Hf}				&		968.2	&	\nuc{228}{Ac}	\\
213.4	&	\nuc{178m}{Hf}				&		1093.5	&	\nuc{208}{Tl} sum	\\
216.7	&		\nuc{178m}{Hf}			&	1120.6		&	\nuc{214}{Bi}	\\
237.4	&		\nuc{178m}{Hf}			&	1238.4		&	\nuc{214}{Bi}  \\
257.6	&		\nuc{178m}{Hf}			&	1242.0		&	\nuc{174}{Lu}\footnote{The weak line at 1241.8 keV most likely originates from Lu-174. It is seen in both the Hf and room background spectra, so it is not associated with the sample. A strong Lu source stored in this radiological controlled area might be the source of this $\gamma$.}		\\
277.3	&		\nuc{178m}{Hf}			&	1377.7		&	\nuc{214}{Bi}	\\
296.8	&		\nuc{178m}{Hf}			&		1460.2	&	\nuc{40}{K}	\\
325.6	&		\nuc{178m}{Hf}			&		1729.6	&	\nuc{214}{Bi}	\\
426.4	&		\nuc{178m}{Hf}			&	1764.8		&	\nuc{214}{Bi}	\\
454.0	&		\nuc{178m}{Hf}			&	1847.4		&	\nuc{214}{Bi}	\\
495.0	&		\nuc{178m}{Hf}			&	2103.6		&	\nuc{208}{Tl}  esc. peak	\\
511.0	&	annihilation				&	2117.8		&	\nuc{214}{Bi}	\\
535.0	&	\nuc{178m}{Hf}				&	2204.1		&	\nuc{214}{Bi}	\\
574.2	&	\nuc{178m}{Hf}				&	2448.4		&	\nuc{214}{Bi}	\\
583.5	&	\nuc{208}{Tl}				&	2614.5		&	\nuc{208}{Tl}  	\\
608.9	&	\nuc{214}{Bi}				&	&   \\

\hline
\end{tabular}
\end{table}

\begin{figure}[t]
\centering
 \includegraphics[width=\columnwidth]{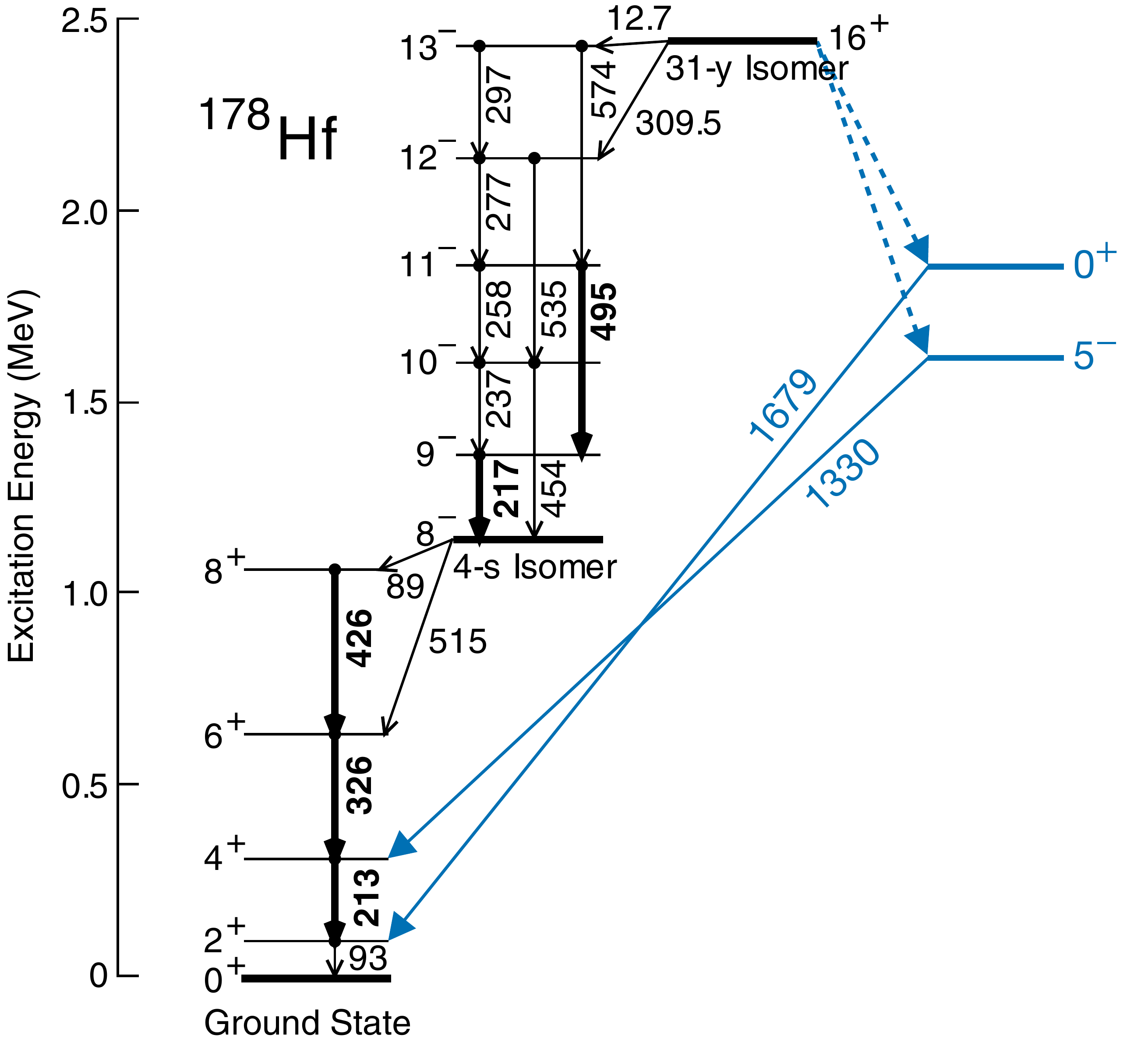}
 \caption{The level diagram of \nuc{178m}{Hf}. To the left (and in black) are the transitions that occur from the SM radioactive decay of the isotope. To the right (and in blue), we illustrate two dark matter-induced transitions of significance: (i) the half-life of the $16^+\to0^+$ transition (associated with the 1679~keV $\gamma$) was the most strongly constrained by this work; (ii) the $16^+\to5^-$ transition (associated with the 1330~keV $\gamma$) provided the strongest constraint on the dark matter-nucleon cross section, $\sigma_n$, among the 11 $\gamma$s considered. This figure was adapted from Fig.\,1 of \cite{Ahmad2001}.}
 \label{fig:LevelDiagraam}
\end{figure}

The metastable state at 2446~keV, with a half-life of 31~yr, has a large spin $J^\pi = 16^+$. Table~\ref{tab:Fsigma} lists a number of Hf states which are prevented from being populated by $\gamma$ transitions originating from the $16^+$ metastable state due to the large spin change, $\Delta J$. An interaction with a heavy and slow dark matter particle, on the other hand, could overcome the $\Delta J$ hindrance and catalyze transitions from the $16^+$ metastable state to those (otherwise unpopulated) lower spin states, which in turn could decay via $\gamma$ emission. Figure~\ref{fig:LevelDiagraam} depicts the \nuc{178m}{Hf} level diagram showing the radioactive decay pathways in contrast to the dark matter-induced pathways for a couple of the key transitions.

When a dark matter particle scatters off the isomeric nuclear state, 
the metastable nuclear energy can be tapped to excite the dark matter to an elevated internal energy state.
The primary advantage of \nuc{178m}{Hf} in our study is its large metastable state energy, which allows us to probe, for the first time in a direct detection experiment, dark matter mass splittings as high as $\delta M_\chi\sim\mathcal{O}(\text{MeV})$. In particular, dark matter-induced transitions of the isomeric state to the lower energy states would be sensitive to the largest dark matter mass splittings. Unfortunately, most of the low-lying \nuc{178}{Hf} states are also populated through the SM radioactive decay of \nuc{178m}{Hf}, and therefore suffer from large backgrounds and result in a significant reduction in sensitivity to an iDM signal. Therefore, we focus on dark matter-induced transitions to higher energy states which are not populated by the SM decay pathways and therefore emit $\gamma$s in regions with reduced backgrounds; these $\gamma$s have energies above $\sim$600 keV. In Table~\ref{tab:Fsigma}, we only list the excited \nuc{178}{Hf} states and associated decay $\gamma$s determined to be the most sensitive given these constraints\footnote{We note that $\gamma$s from the dark matter-induced states at 1731.1~keV and at 1781.3~keV are also ignored because they fall in high background regions.}. These states balance the need for high dark matter-induced transition energy against the backgrounds at the $\gamma$ energy.

For the candidate $\gamma$s  listed in Table~\ref{tab:Fsigma}, we obtain the number of observed and expected background counts within an optimized window centered on the $\gamma$ energy. The expected background count (B) is obtained by performing a side-band fit of the data, assuming a linear background distribution. The width of the optimal window is deduced by maximizing $\mathcal{A}_\gamma/\sqrt{B}$, where $\mathcal{A}_\gamma$ is the signal acceptance within the chosen window, assuming it follows a gaussian distribution. For a continuum spectrum, the resulting signal acceptance within the optimal window is given by $\mathcal{A}_\gamma\simeq0.84$; however, since we collected a binned spectrum, the signal acceptance for each $\gamma$ received a small correction to account for the bin edges. There were no observable peaks at any of the smoking-gun energies in Table~\ref{tab:Fsigma}.

\begin{table*}[htp]
\caption{The input parameters\footnote{The level energies, transition energies, and branching ratios are taken from the National Nuclear Data Center NuDat database \cite{ACHTERBERG2009} accessed March 2023.} and resulting 90\% C.L. limits on the half-life $T_{1/2}^{(j)}$ of the dark matter-induced transition shown in eq.\,(\ref{transitionProcess}). The smoking gun signal for this process is the $\gamma$ emitted in the decay of dark matter induced state $^{178}\text{Hf}_j$ (4th column). The relative detector efficiencies normalized to that for the 495-keV $\gamma$, $\epsilon^{(j)}_{\gamma}/\epsilon_{495}$, were obtained from \cite{ORTEC} and are assumed to carry uncorrelated uncertainties of $\pm20\%$. The isomeric state of \nuc{178m}{Hf} at 2446.1 keV has $J^{\pi}= 16^+$ and $K=16$.
} 
\label{tab:Fsigma}
\centering
\begin{tabular}{ccccccccrc}
\hline
label~ &	state energy	& state &  $\gamma$ energy 	& $\gamma$ branching~	&  ~acceptance~ & rel.\,eff.& background &observed~~& 	 $T_{1/2}^{(j)}$ \\
$j$& ~$E_{j}$(keV)~  & ~~$J^{\pi},\,K$~~~ & $E^{(j)}_\gamma$(keV) & ratio $b^{(j)}_\gamma$ & $\mathcal{A}^{(j)}_{\gamma}$ &   $\epsilon^{(j)}_{\gamma}/\epsilon_{495}$ &counts& counts~\,\,&($10^{5}$~yrs)
\\
\hline\hline
1&	1635.6	& 	$4^+$,~0	&	1542.2	&~~$0.97\pm0.04$~~	&	0.83	&	0.41		&	$20.52\pm1.29$	&	17~~~~	&	$>1.56$\\
2&	1636.7	&      $5^-$,~5	&	1330.0	&	$0.57\pm0.03$		&	0.86	&	0.44	&	$44.23\pm1.98$	&	32~~~~	&	$>1.12$\\
3&	1640.5	&	$5^+$,~4	&	1333.8	&	$0.55\pm0.02$		&	0.86	&	0.44	&	$40.93\pm1.87$	&	40~~~~	&	$>0.52$\\
4&	1648.8	&	$6^-$,~2	&	1016.6	&	$0.60\pm0.04$		&	0.84	&	0.47		&	$75.97\pm2.52$	&	82~~~~	&	$>0.30$\\
5&	1651.5	&	$5^-$,~1	&	1344.9	&	$0.70\pm0.02$		&	0.86	&	0.44	&	$38.27\pm1.81$	&	41~~~~	&	$>0.51$\\
6&	1654.3	&	$4^+$,~0	&	1348.0	&	$0.68\pm0.21$		&	0.86	&	0.44	&	$36.99\pm1.81$	&	34~~~~	&	$>0.69$\\
7&	1691.1	&	$6^+$,~2	&	1059.0	&	$0.55\pm0.02$		&	0.83	&	0.47		&	$74.65\pm2.45$	&	50~~~~	&	$>1.21$\\
8&	1697.5	&	$9^-$,~8	&	333.4	&	$1.00\pm0.00$		&	0.83	&	1.24		&	$6434.7\pm29.8$~\,	&	6506~~~~~&	$>0.14$\\
9&	1747.1	&	$4^-$,~2	&	1440.6	&	$0.12\pm0.03$		&	0.84	&	0.44		&	$33.21\pm0.84$	&	31~~~~	&	$>0.13$\\
10&	1772.1	&	$0^+$,~0	&	1678.8	&	$1.00\pm0.11$		&	0.81	&	0.25		&	$15.55\pm0.81$	&	11~~~~	&	$>1.79$\\
11&	1788.6	&	$6^+$,~4	&	1156.3	&	$0.45\pm0.04$		&	0.81	&	0.47		&	$63.84\pm0.81$	&	67~~~~	&	$>0.26$\\
\hline
\end{tabular}
\end{table*}

To establish our notation, consider the process in which a dark matter particle upscatters off the isomeric nuclear state \nuc{178m}{Hf}, causing the nucleus to transition to a lower energy level $^{178}\text{Hf}_j$,
\begin{equation}\label{transitionProcess}
\chi+{^{178m}\text{Hf}}~\to~ \chi^*+{^{178}\text{Hf}_j}.
\end{equation}
We will denote the inelastic cross section for this process by $\sigma_{inel}^{(j)}$. The produced state $^{178}\text{Hf}_j$ can then de-excite via emission of a $\gamma$ of energy $E_\gamma^{(j)}$  with a branching ratio given by $b_{\gamma}^{(j)}$.
The expected signal count for the process described above, $S^{(j)}_\gamma$, is given by
\begin{equation}\label{darkmattergammaRate}
S^{(j)}_\gamma = N_T\Delta t \times(\sigma_{inel}^{(j)}\,\Phi_\chi)\times(b_{\gamma}^{(j)}\mathcal{A}^{(j)}_{\gamma}  \epsilon_{\gamma}^{(j)}),
\end{equation}
where $N_T$ is the number of target \nuc{178m}{Hf} nuclei; $\Delta t=974.2$~s is the live time; $\Phi_{\chi}$ is the dark matter flux; and $\mathcal{A}^{(j)}_{\gamma}$ and $\epsilon_\gamma^{(j)}$ are, respectively, the signal acceptance within the region of interest and detection efficiency for the $\gamma$ emitted in the decay of $^{178}\text{Hf}_j$. 

The number of target atoms $N_T$ can be deduced from the SM activity of the Hf sample. In particular, the SM decay chain of \nuc{178m}{Hf} produces a 495-keV $\gamma$ line with a probability $p_{495}=0.736\pm0.014$ \cite{PhysRevC.68.031302}. The number of 495-keV $\gamma$ counts observed during our live time, $S_{495}=96,808\pm395$, relates to $N_T$ via:
\begin{equation}\label{nHfatoms}
N_T = \frac{\tau_\text{isomer}}{\Delta t}\,\frac{S_{495}}{\,p_{495} \,\epsilon_{495}\,},
\end{equation}
where $\tau_\text{isomer}=1.41\times10^{9}$\,s is the lifetime of \nuc{178m}{Hf} and $\epsilon_{495}$ is the detection efficiency for the 495 keV $\gamma$ line. 

Combining (\ref{darkmattergammaRate}) and (\ref{nHfatoms}), we can express the dark matter event rate (i.e., number of scattering events in (\ref{transitionProcess}) per unit time per target nucleus) as
\begin{equation}\label{PhiSigma}
\sigma_{inel}^{(j)}\,\Phi_\chi ~=~ \tau_\text{isomer}^{-1}\, \frac{S^{(j)}_\gamma}{S_{495}}\, \frac{p_{495}\, \epsilon_{495}}{b_{\gamma}^{(j)}\mathcal{A}^{(j)}_{\gamma}\, \epsilon_{\gamma}^{(j)}}.
\end{equation}
The corresponding half-life for this dark matter-induced transition is simply related to (\ref{PhiSigma}) via
\begin{equation}
T_{1/2}^{(j)}=\frac{\log{2}}{\sigma_{inel}^{(j)}\,\Phi_\chi}.
\end{equation}

By performing a profiled log-likelihood fit of the signal strength for each of the 11 $\gamma$ lines considered, we obtained 90\% confidence level (C.L.) limits on the half-lives $T_{1/2}^{(j)}$, given in Table~\ref{tab:Fsigma}.

The inelastic cross section for the process in (\ref{transitionProcess}), $\sigma_{inel}^{(j)}$, can be related to the model-dependent dark matter-nucleon cross section $\sigma_n$ using the formalism of \cite{Pospelov2020}. We used this relation to translate our limits into constraints on the parameter space of inelastic dark matter (iDM), namely, $\sigma_n$ versus the iDM mass splitting $\delta M_{\chi}$ for a benchmark iDM mass of $M_\chi=1$~TeV. Note that the transition with the most strongly constrained half-life does not necessarily provide the strongest constraint on $\sigma_n$. That is because the nuclear form factor, which suppresses the transition rate in (\ref{transitionProcess}), depends not only on the momentum transfer $q$ and change in angular momentum $\Delta J$, but also on $K$-selection rules. Specifically, each nuclear state has a $K$-quantum number given by the projection of its angular momentum on its symmetry axis (see Table~\ref{tab:Fsigma}), and transitions with $\Delta K$ greater than the multipolarity of the emitted radiation suffer from an additional suppression, the so-called ``$K$-hindrance''. Among the 11 transitions considered, $j=2$ provides the strongest constraint on $\sigma_n$, since it has the second smallest $\Delta K(=\!11)$, and the observed counts for its associated $\gamma$ line of 1330~keV showed a $\sim$2$\sigma$ deficit relative to the background expectation. We can contrast the constraining power of $j=2$ with that of other transitions. For example, while $j=8$ has the smallest $\Delta K(=\!8)$, its associated 333.4~keV $\gamma$ line lies in a region with substantial backgrounds ($\sim2$ orders of magnitude larger than the backgrounds for the other lines), which significantly weakens its dark matter constraining power. As another example, transitions $j=1$ and $j=10$ have the most strongly constrained half-lives, but also the largest $\Delta K(=\!16)$, which suppresses their rate for $\delta M_\chi\gtrsim 600$ keV. For further technical details, we refer the reader to \cite{Pospelov2020}.

\begin{figure}[t]
 \centering
 \includegraphics[width=\columnwidth]{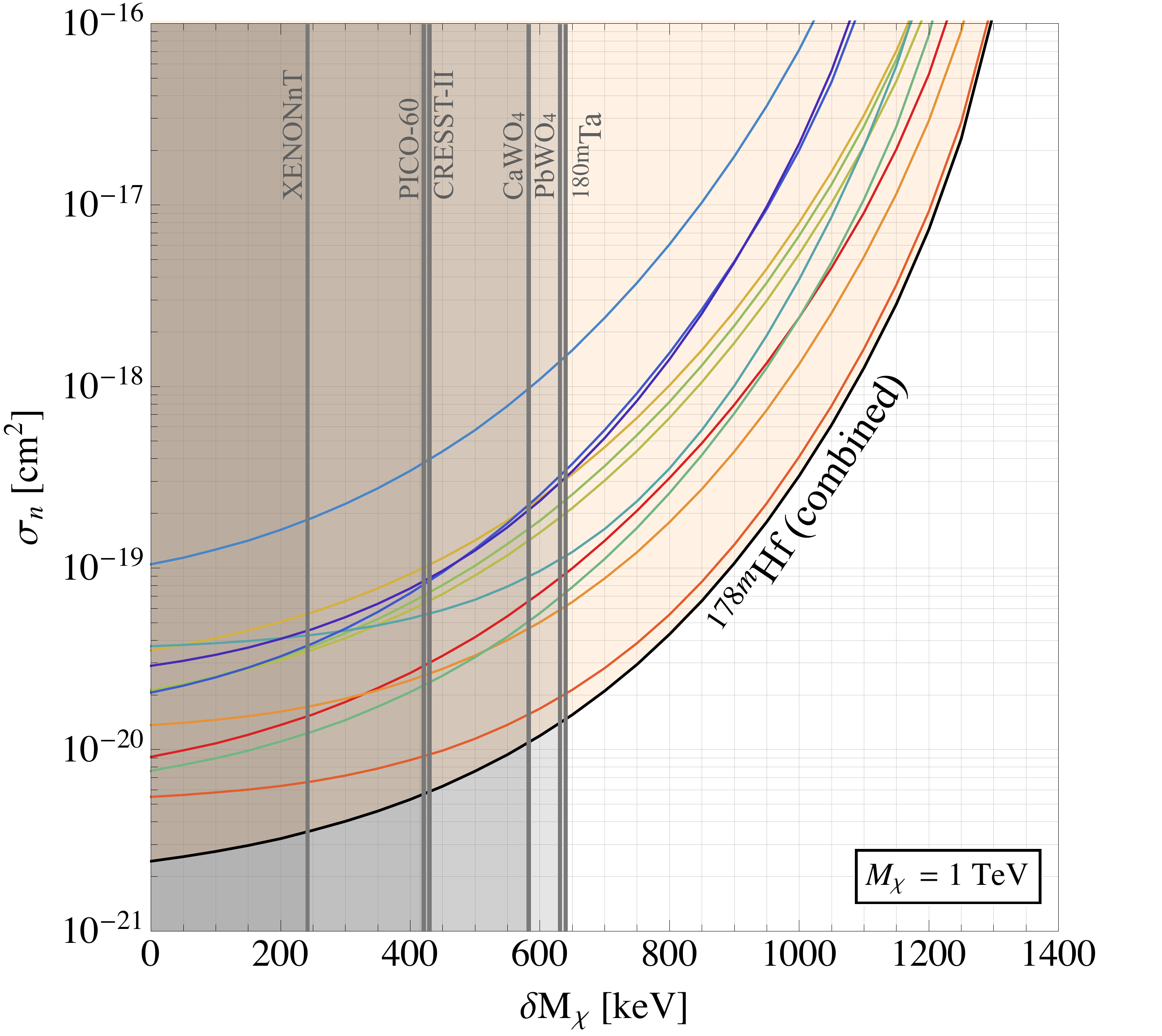}
 \caption{The 90\% C.L. exclusion limits on the parameter space of inelastic dark matter, assuming a standard halo model with local dark matter density $\rho_\chi=0.3\,\text{GeV/cm}^3$ and galactic escape velocity $v_\text{esc}=600$ km/s. The black curve shows the combined limit from all 11 $\gamma$s, and the shaded gray regions show previous existing limits \cite{Song2021,PhysRevLett.124.181802,Beeman2013,Munster2014,Angloher2016,PhysRevD.93.052014,PhysRevLett.121.111302}. The 11 colored curves show the limits from each individual $\gamma$ line in Table \ref{tab:Fsigma}. Following the $j$-label convention of Table \ref{tab:Fsigma}, and fixing $\delta M_\chi=700\text{ keV}$, the order of the 11 colored curves, from stronger to weaker exclusion, is: $j=2,\,3,\,7,\,1,\,8,\,5,\,6,\,4,\,11,\,10,\,9$.}
 \label{fig:Limits}
\end{figure}

Finally, we performed a joint log-likelihood fit of the signal strength for all the 11 $\gamma$s combined. Our results are shown in Fig.\,\ref{fig:Limits}.

For $\delta M_{\chi}\gtrsim 640$~keV, our experimental limit on $\sigma_n$ is the best to date, albeit dark matter models with such large cross sections would necessarily come from composite dynamics and might face model building challenges. The strongest competing constraint in the large mass splitting region $\delta M_{\chi}\gtrsim400$~keV comes from searches using the tantalum metastable state \nuc{180m}{Ta}, which has a very long half-life and is naturally abundant albeit with a low isotopic fraction. This has enabled experiments with samples containing a large number of  \nuc{180m}{Ta} nuclei, resulting in interesting limits on SIDM and iDM models~\cite{PhysRevC.95.044306,PhysRevLett.124.181802,cerroni2023deepunderground}, as well as ongoing experiments with significantly improved reach~\cite{arnquist2023constraints}. Still, the comparably lower metastable energy of  \nuc{180m}{Ta} (76.3~keV, in contrast to 2446~keV for \nuc{178m}{Hf}) limits its sensitivity to $\delta M_{\chi}\lesssim600$~keV.
Other existing experimental results are described in \cite{Song2021}, with specific limits derived from data for PbWO$_4$~\cite{Beeman2013}, CaWO$_4$~\cite{Munster2014}, CRESST-II~\cite{Angloher2016}, PICO-60~\cite{PhysRevD.93.052014}, and XENONnT~\cite{PhysRevLett.121.111302}.

A number of improvements and options for future measurements are possible. For a repeat of the measurements presented here, a longer run time and the use of a shielded detector could improve sensitivity by a factor of $\sim$10. A high-efficiency Ge-detector array similar to AGATA~\cite{BRACCO2021}---with a large solid angle acceptance and detectors distant enough from the source so as to have a manageable rate---could further improve sensitivity by an additional factor of $\sim$100.

The Hf measurements reported here were performed at a surface site, and therefore were not sensitive to viable parameter space in SIDM. By deploying the Hf sample and a detector underground and repeating these measurements, perhaps at multiple depths, one could probe the effects of a dark matter traffic jam~\cite{Pospelov2020} in models of SIDM (both elastic and inelastic).

Ideally, one would like a very large sample of the Hf isomer. For this Hf sample, 1~kg of Ta was irradiated for 60 days with an 800 MeV, 350~$\mu$A proton beam. The Ta target was used as a dedicated beam stop from which the Hf was extracted by radiochemistry~\cite{Taylor1998}. If feasible, processing additional targets could produce a large quantity of \nuc{178m}{Hf}; however, the cost of the required radiochemistry would have to be weighed against the science reach. Furthermore, practical difficulties associated with the high radioactivity of the sample would need to be overcome, such as a fast, highly efficient detection system in order to fully exploit the science scope of such measurements.

An alternative experimental setup could be arranged to search for the decay of the upscattered dark matter particle. Specifically, dark matter particles could inelastically scatter off \nuc{178m}{Hf} to an excited state,  and decay promptly or with $\mathcal{O}$(m) displacements via emission of a monochromatic $\gamma$, which might be observed in a nearby detector (see, e.g., \cite{Pospelov2014}.)
For this geometry, a large sample of \nuc{178m}{Hf} (such as beam stops that have been irradiated for extended periods) could be located some distance from a detector. A Ge detector could be sited nearby to search for such an anomalous $\gamma$. While number of old tungsten and tantalum-cladded tungsten beam stops are stored on location behind shielding but our early assessment is that none have enough \nuc{178m}{Hf} for a useful measurement. A dedicated production would be required.

\noindent{\it Acknowledgements}~~We thank Evelyn Bond and Athena Marie Marenco for assisting with access to the sample. We gratefully acknowledge support from the U.S.~Department of Energy Office of Science, and from the Los Alamos National Laboratory's Directed Research and Development (LDRD) Program for this work.

\bibliography{Hfanalysis}

\end{document}